\documentclass[12pt]{article}
\usepackage{graphicx}
\usepackage{amssymb, amsmath}
\usepackage{natbib}
\usepackage{hyperref}
\usepackage{xcolor}
\usepackage[normalem]{ulem}
\usepackage[acronym]{glossaries}
\providecommand{\keywords}[1]
{
  \small	
  \textbf{\textit{Keywords---}} #1
}

\title{Dissent and Rebellion in the House of Commons: A Social Network Analysis of Brexit-Related Divisions in the 57$^{ th}$ Parliament}

\author{Carla Intal$^1$ and Taha Yasseri$^{1,2,3,4}$\footnote{Corresponding author. Email: taha.yasseri@ucd.ie }\\
{\small $^1$Oxford Internet Institute, University of Oxford, Oxford, UK} \\ 
{\small $^2$School of Sociology, University College Dublin, Dublin, Ireland}\\
{\small $^3$Geary Institute for Public Policy, University College Dublin, Dublin, Ireland}\\
{\small $^4$Alan Turing Institute, London, UK}\\}

\begin{document}
\maketitle
\begin{abstract}
The British party system is known for its discipline and cohesion, but it remains wedged on one issue: European integration. We offer a methodology using social network analysis that considers the individual interactions of MPs in the voting process. Using public Parliamentary records, we scraped votes of individual MPs in the 57th parliament (June 2017 to April 2019), computed pairwise similarity scores and calculated rebellion metrics based on eigenvector centralities. Comparing the networks of Brexit- and non-Brexit divisions, our methodology was able to detect a significant difference in eurosceptic behaviour for the former, and using a rebellion metric we predicted how MPs would vote in a forthcoming Brexit deal with over 90\% accuracy.

\end{abstract}

\vspace{1cm}

\section*{Background}

The British party system is widely known to be a strong and disciplined one. Throughout its contemporary history, its two main parties\textemdash Labour and Conservative\textemdash has lent credibility to the Parliamentary process, setting the landscape for the effective implementation of policies in the British government.  

It is notable, however, that the cohesion and unity in the modern British party system is persistently wedged by one issue, which is that of European integration. It has been observed in history at least twice - first, in the 1970s as the UK elected to join the European Economic Commission (EEC); and second, in the 1990s as then-Prime Minister Margaret Thatcher famously retracted her support to the priorities and direction of the European Union (EU). The dissent and rebellion by the British parliament around European integration (`euroscepticism') has long been studied by political scientists and historians and is well-documented in qualitative work \citep{wood_comparing_1982, baker1999backbench, whiteley1999discipline, cowley_rebels_1999, forster_euroscepticism_2002}.  Meanwhile, empirical studies often cite the Rice score as a standard measure for cohesion and conformity in politics \citep{rice1938, norton_changing_1980, collie_voting_1984, garner_party_2005, sieberer_party_2006, tzelgov_cross-cutting_2014}.  This is a metric between zero and one, and is a ratio comparing those in favour of an issue versus those who are not.  And while it is widely used in quantitative work, the metric has a few limitations, primarily, it is a simplistic measure that does not consider more than two voting options (whereas the British parliament could vote either `Aye', `No' or abstain); second, it is an aggregate measure reported at the party level and does not measure individual cohesion in relation to a group; and (thus) third, the metric is not directly sensitive to shifts in pairwise interactions amongst members of parliament (MPs) over a voting period.  But voting is a social phenomenon and it is worth considering that an MP's voting record may change over time during the course of a legislature, whether in relation to another MP, or more widely, in relation to a group.  Similarly, other empirical analyses on party cohesion and rebellion \citep{searing_measuring_1978, collie_voting_1984, pattie1998, russell_parliamentary_2014, raymond_explaining_2017, surridge_brexit_2018} also cite differences at an aggregate level.

We aim to address these limitations by using social network analysis.  This paper seeks to investigate whether methods in social network analysis can help detect the continuing `eurosceptic' behaviour of the Parliament in the Brexit era. We particularly analyse the voting patterns of Members of the 57th Parliament on Brexit-related matters to identify cohesion and/or rebellion.  We define rebellion among \textit{pairs} of MPs as one of two forms: first, as a \textbf{cross-party alliance}, where an MP defies his/her own party whip and votes in coherence with another MP from a different party; and second, as a \textbf{within-party conflict}, where two MPs belong to the same party but vote differently. 

The primary contribution of this methodology is to inform voting cohesion (or dissent) at the individual level rather than report them in aggregate at the party- or group-level.  Second, this analysis on individual choice is advantageous in that it can help anticipate voting patterns by MPs as they unfold, compared to a ex-post retrospective analysis using historical accounts.

\section*{Data and Methods}

\subsection*{Data}

Using data from Hansard, the official central repository of all UK Parliamentary records, we begin the analysis with a dataset of divisions in the House of Commons for only the 57th Parliament, which is the legislature immediately following the Brexit referendum. Our analysis period is from 21$^{\rm{st}}$ June, 2017 until 10$^{\rm{th}}$ April, a total of 414 divisions (A division is a vote of either ``Aye (Yes)'' or ``No'' on a Parliamentary issue). Hansard provides comprehensive and verbatim information on Parliament debates, divisions, petitions, and statements, for both the House of Commons and the House of Lords.  The digitalised database of division voting started in March 2016, and is made available for download through the data.parliament.uk website.   Committee divisions, which tackle particular areas of interest, were excluded from the analysis.

Hansard collects data for each division in a CSV file, tagged with a unique ID, a unique URL which contains detailed information (such as the voting record), and the title of the division. We identify the 414 that belong to the 57th Parliament, and manually classified each division to a \textit{Brexit} or \textit{non-Brexit} group for the analysis. For instance, any division that mentions the words: EU exit, EU withdrawal, Brexit, and other related keywords in their title were tagged as \textit{Brexit} divisions. Out of the 414, there were 192 divisions labeled as \textit{Brexit} and 222 \textit{non-Brexit}. Then, to gather the names of every MPs who voted ``Aye'' or ``No'' on each of the 414 divisions, we built a web crawler that would loop across the unique URLs for each of the 414 divisions, each page containing a JSON file with information on each MP's vote. 

MPs either belong to a political party or are Independent. The breakdown of 650 MPs by political party are: 317 Conservative, 254 Labour, 35 SNP or the Scottish National Party, 11 Liberal Democrats, 10 DUP or Democratic Unionist Party (of Northern Ireland), 7 Sinn F\'ein, 4 Plaid Cymru (or the Party of Wales), 1 Green Party, and 11 Independent.  Out of the 650, two from the Conservative and two from the Labour party do not vote as the Speaker and 3 Deputy Speakers of the House, while the 7 Sinn F\'ein are absentionists and do not take their seats in the House of Commons; hence there are 639 voting MPs. Because Hansard only records the MPs who voted ``Ayes'' or ``Noes,'' MPs that were missing from either were presumed to not have voted on the division.  This is the main dataset for the analysis.

We also label each MP by left-wing or right-wing, depending on their political party affiliation. Traditionally, the two largest parties, Labour and Conservative, are on the left and the right, respectively.  The DUP, which entered into a coalition with the Conservative Party in 2017, is also on the Right-wing; while the other Opposition parties are on the Left-wing. (The Liberal Democrats, which at times are aligned with the Conservative party, were classified as left-wing for this analysis because of their Remain stance on Brexit.) Meanwhile, for each of the 11 Independent MPs who, by definition is not affiliated, we labelled their ideology by their prior political affiliation: one Independent MP was formerly Conservative, while 8 were formerly Labour, and one was a former Liberal Democrat.  One Independent candidate (Lady Hermon) was classified as Left-wing, primarily because of her refusal to support the right-leaning Conservative Party \citep{bbc_mp_2010} to which her former party\textemdash the Ulster Unionist Party (UUP)\textemdash is associated with. The political party breakdown of the 639 voting MPs is provided in Table \ref{t1}.  

\begin{table}[htbp]
\centering
\begin{tabular}{lrr}

\textbf{Political Party} & \textbf{N} & \textbf{\begin{tabular}[c]{@{}c@{}}Left or\\ Right-wing\end{tabular}} \\ \hline
Conservative & 315 & Right \\
Labour & 252 & Left \\
Scottish National Party & 35 & Left \\
Independent & 11 & 10 Left, 1 Right \\
Liberal Democrat & 11 & Left \\
Democratic Unionist Party & 10 & Right \\
Plaid Cymru & 4 & Left \\
Green Party & 1 & Left \\ \hline
Total & \multicolumn{1}{l}{639} & \multicolumn{1}{l}{\textbf{}} \\ 
\end{tabular}
\caption{Overview of the MPs under study and their party affiliations.}
\label{t1}
\end{table}

\subsection*{Methods}
\subsubsection*{Cosine Similarity}
We start with a 639 $\times$ 414 matrix.  On the rows are each of the 639 voting MPs, and on the columns are each of the 414 divisions.  The elements of the matrix are: +1, if the member voted ``Aye'' on the division, $-$1 if the member voted ``No'', and 0 if the MP did not vote.

We then sub-divided the matrix into the \textit{Brexit} and the \textit{non-Brexit} divisions.  As mentioned earlier, these were classified manually according to the mention of EU exit, withdrawal, or related words in their division title. Splitting the big matrix results in two smaller matrices with sizes 639 $\times$ 192 containing the Brexit divisions, and 639 $\times$ 222 containing the non-Brexit divisions.

Using the cosine similarity formula, which projects a pair of vectors in multidimensional space, we transform the original matrix to a square matrix with dimension 639 x 639, to project the similarity among each vector pair of MPs, controlling for the differences in the number of non-zero elements in the vectors:

\begin{equation}
\label{eq2}
{\frac{\mathbf {a} \cdot \mathbf {b}}{\|\mathbf {a} \|\|\mathbf {b} \|}}={\frac {\sum \limits _{i=1}^{n}{a_{i}b_{i}}}{{\sqrt {\sum \limits _{i=1}^{n}{a_{i}^{2}}}}{\sqrt {\sum \limits _{i=1}^{n}{b_{i}^{2}}}}}}
\end{equation}
$$ \text{where }a_i \text{ and } b_i \text{ are vectors of votes for division }i \text{ for a pair of MPs } \mathbf {a} \text{ and } \mathbf {b}, $$
$$
\text{and } n = 
\begin{cases}
			192, & \text{if \textit{Brexit}}\\
            222, & \textit{non-Brexit}
		 \end{cases}
$$

\bigskip
Each vector pair (i.e., pair of MPs) generated from the matrices in equation (\ref{eq2})  will contain elements ranging from between $-$1 (full dissimilarity between the voting pair) and +1 (full similarity). A value of 0 means no correlation.

Note that the cosine similarity formula considers varying vector densities, which means that the voting similarity score is weighted by the frequency of the votes cast. Thus, the differences in frequencies may represent MPs who abstain or do not vote on certain divisions.

\subsubsection*{Party and ideology similarities}

The cosine similarity matrices in equation (\ref{eq2}) only establishes the voting coherence or voting polarity, but does not indicate anything on the party similarities between pairs of MPs. Hence, we define a reference party similarity matrix and compare the cosine similarity matrices to it.

Consider a 639 $\times$ 639 matrix with entries of +1, if the MP pair belongs to the same political party; or $-$1, if the MP pair does not belong to the same party. Note that there are 8 parties (excluding Sinn F\'ein) in the analysis: Conservative, Labour, Liberal Democrats, DUP, SNP, Plaid Cymru, Green Party, and Independent; though this matrix is also prepared for a two-ideology case (Left-wing and Right-wing). The party similarity matrix is identical for the Brexit and the non-Brexit case. 

\subsubsection*{Party and ideology-adjusted voting similarities}

Finally, we compare the cosine similarity matrix with the party similarity matrices by subtracting the latter from the former for both Brexit and non-Brexit divisions.  

The elements of either matrix have a range of values from $-$2 to +2, centred on zero.  These values indicate the magnitude by which the voting similarity between a pair of MPs is explained by their party.  For instance, a value that is equal to zero means that the voting similarity between MPs is fully matched to what is expected from their party affiliations.  As values move further away from zero, this means that the party affiliations have less power in explaining the voting similarity/dissimilarity among pairs of MPs. The same methodology and interpretation is applied to the 2-ideology case.

Thus, the non-zero values in the matrix imply varying deviations from voting patterns expected by the party affiliations.  Values that are near-zero, for instance, mean that voting (dis)similarity is \textemdash for the most part\textemdash explained by the party (dis)similarity.  But of interest to this study are the cases where the MPs' voting (dis)similarity is \textbf{not} explained by their party affiliations, or the values that deviate the farthest away from zero, towards the end of the range [$-$2, +2]. 
Values close to $-$2 is an indication of a \textbf{within-party conflict}, or a pair of MPs that voted dissimilarly despite belonging to the same party, while values close to +2 is an indication of a \textbf{cross-party alliance}, or a pair of MPs that voted cohesively despite belonging to different parties.

\subsubsection*{Network representation}
Finally, the result is a square matrix with a zero main diagonal and symmetric entries, which is equivalent to the adjacency matrix of an undirected graph.  The row and column labels (which are the names of the MPs) are the nodes of the network, while the elements of the matrix are the edges \textemdash they refer to the party-adjusted voting similarity or dissimilarity of each pair of nodes (or pair of MPs). 

One can then visualise the adjacency matrix into a network, where each node represents an MP, and the elements of the adjacency matrix suggests the relationship between node pairs: strong negative values indicate polar opposites (or within-party conflict), while strong positive values mean cohesion (or cross-party alliance). Using a force-directed layout, pairs of MPs that correspond to a high negative score (within-party conflict) in the adjacency matrix are pulled further apart in the network space, while pairs of MPs with a high positive score (cross-party alliance) are expected to be pushed closer together. The Supplementary Material provides more details.

\subsubsection*{Eigenvector centrality}
To investigate the cohesion and divergence between MPs we will first use visualization of the network representation of the votes cast by them. However, to have a more systematic and reproducible approach, beyond visual inspection, we continue by identifying the outliers through calculating the eigenvector centrality of the MPs within the network. There are several propositions for calculation of centrality in a signed network (see \cite{bonacich2007some,tang2016survey} for examples). Some consider the sign of the connections in the calculation directly and some split the network into positive and negative sub-networks and calculate the centrality separately. We take the latter approach and calculate the eigenvector centrality of each node once only considering positive edges, and then only negative edges where the weights are replaced with the absolute values. The final centrality score of each MP is the sum of the two centrality scores. 

\section*{Results}

\subsection*{Party and ideology-adjusted voting similarity matrices}
\begin{figure}[htbp]
\centering
\includegraphics[width=\textwidth]{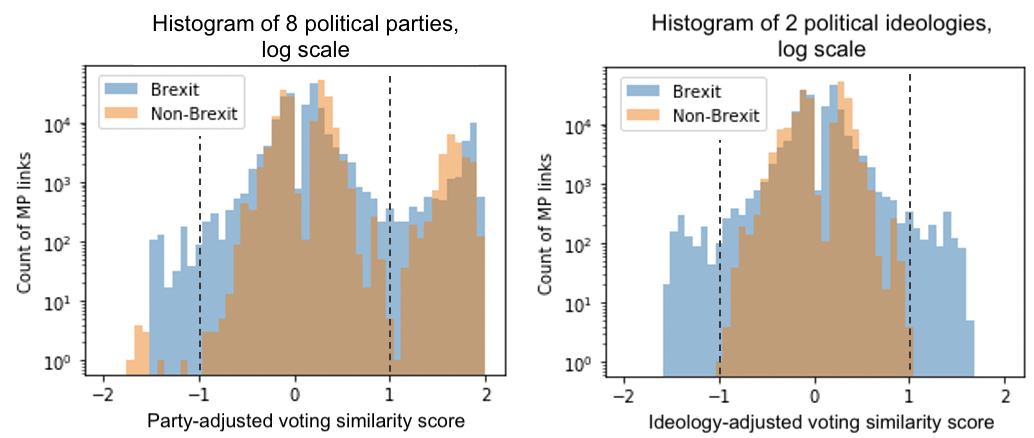}
\caption{Histogram of voting similarity scores comparing Brexit and non-Brexit divisions. Left: 8 political parties; Right: 2 ideologies.  Values that lie beyond the dotted line close to the extremes represent connections between dissenting MPs.}
\label{8pp}
\end{figure}

In Figure \ref{8pp}, we show a histogram of the links of the matrices from the party-adjusted voting similarity matrix. For both the 8-party case and the 2-idealogy case, the Kolmogorov-Smirnov (KS) test confirms that the two distributions (Brexit vs. non-Brexit) are different with an alpha of $<$0.001.

The two types of dissent are presented in the histogram as follows.  Values that are greater than or equal to 1 are \textbf{cross-party alliances}, or two MPs that belong to different parties but voted cohesively.  On the other extreme, values that are less than or equal to $-$1 signal \textbf{within-party conflicts}, or two MPs that belong to the same party but voted opposite. There are two things that can be inferred from Figure \ref{8pp}: first, there are comparatively higher levels of cross-party alliances than within-party conflicts; second, and more importantly, it can be observed that there is a larger frequency of within-party conflicts on Brexit-related divisions compared to non-Brexit, which is consistent with the literature on party rebellion particularly on issues of European integration.

\subsection*{Network projection: 8-party case}

\begin{figure}[htbp]
\centering
\includegraphics[width=\textwidth]{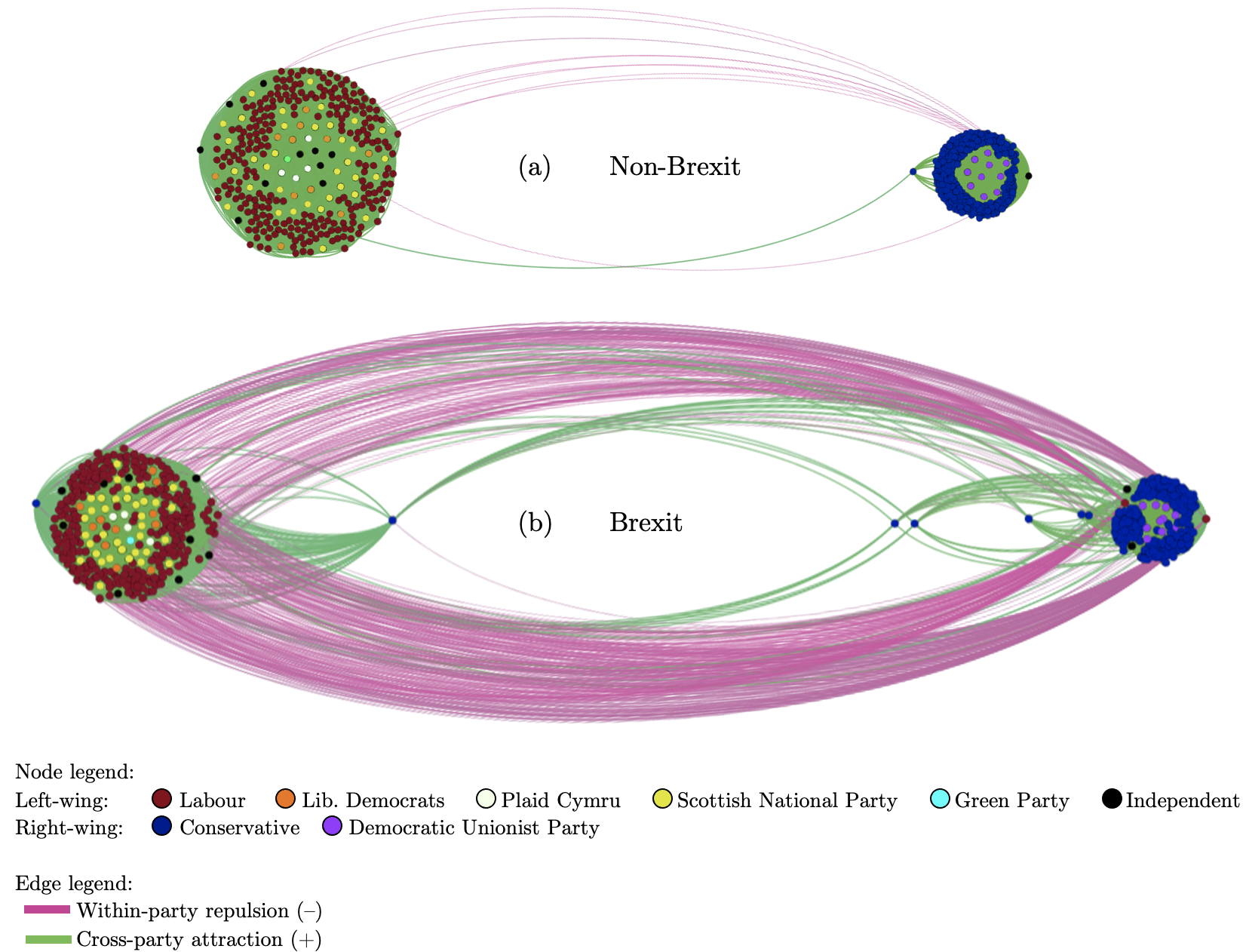}
\caption{Network projection of dissent and rebellion for non-Brexit (a) and Brexit (b) divisions, for the 8 political party case. Each node denotes an MP, connected by an edge to another MP.  The colour of the edge represents whether the connection is a repulsive (push) or attractive (pull) force.  Node colours denote party affiliation.}
\label{8p_network}
\end{figure}

We visualise the extreme values, i.e. $\geq1$ (cross-party alliances) and $\leq-1$ (within-party conflicts) in Figure \ref{8p_network}. 
It can be observed that for the top graph, which represents the non-Brexit divisions, there are two distinct clusters, one largely blue, and the other is largely red; and the nodes that reside in each cluster are connected by dense green edges.  Meanwhile, there are very few cross-cluster connections (mostly related to Liberal Democrat MPs). 

The non-Brexit graph suggests strong party discipline within each of the two main parties, Conservative and Labour.  Conservative nodes tend to cluster together, while Labour nodes (and other parties in opposition) are in the other cluster.  This means that for non-Brexit divisions, MPs vote largely within their party lines, and very rarely do they rebel and cross to the other cluster.  One can also infer the cross-party alliances from the graph.  For instance, the DUP and the Conservative are clustered together, which verifies their voting coalition.  Meanwhile, the Labour, Plaid Cymru, Green, and SNP, are on the other cluster, which suggests a united opposition when voting in Parliament on non-Brexit divisions; and Liberal Democrats split between the two camps. 

This is a stark contrast to the Brexit divisions, the bottom graph of Figure \ref{8p_network}.  There is evidently greater cross-cluster interaction, which is an indication that party lines are blurred.  The rose-coloured edges suggest within-party conflict.  This means that there are a number of MPs that rebel from the cluster discipline and are pulled apart. For example, some nodes representing Labour MPs have moved within the cluster space of the Conservative party, which indicates that these MPs have voted more cohesively with the Conservative rhetoric, and thus expressing strong repulsion to MPs from their own party.  Also, cross-party alliances may occur among pairs of ideologically polarised MPs; several nodes representing the Conservative MPs are being pushed to the centre towards the Labour cluster.

These results are however largely influenced by the number of political parties in the analysis.  Mainly, the cross-party alliances occur almost exclusively within the ideological cluster, and does not necessarily signify rebellion.  In order to distinguish the true rebellions from the effect of ideological cohesion from multiple parties, we re-label the party similarity matrix to indicate ideological affiliation (Left- or Right-wing) rather than the original 8-political party classification.  The analysis based on the 2-ideology case is also backed by recent studies that find ideology as a determinant of Brexit support \citep{vasilopoulou_uk_2016, surridge_brexit_2018}.

\subsection*{Network projection: 2-ideology case}
\begin{figure}[htbp]
\centering
\includegraphics[width=\textwidth]{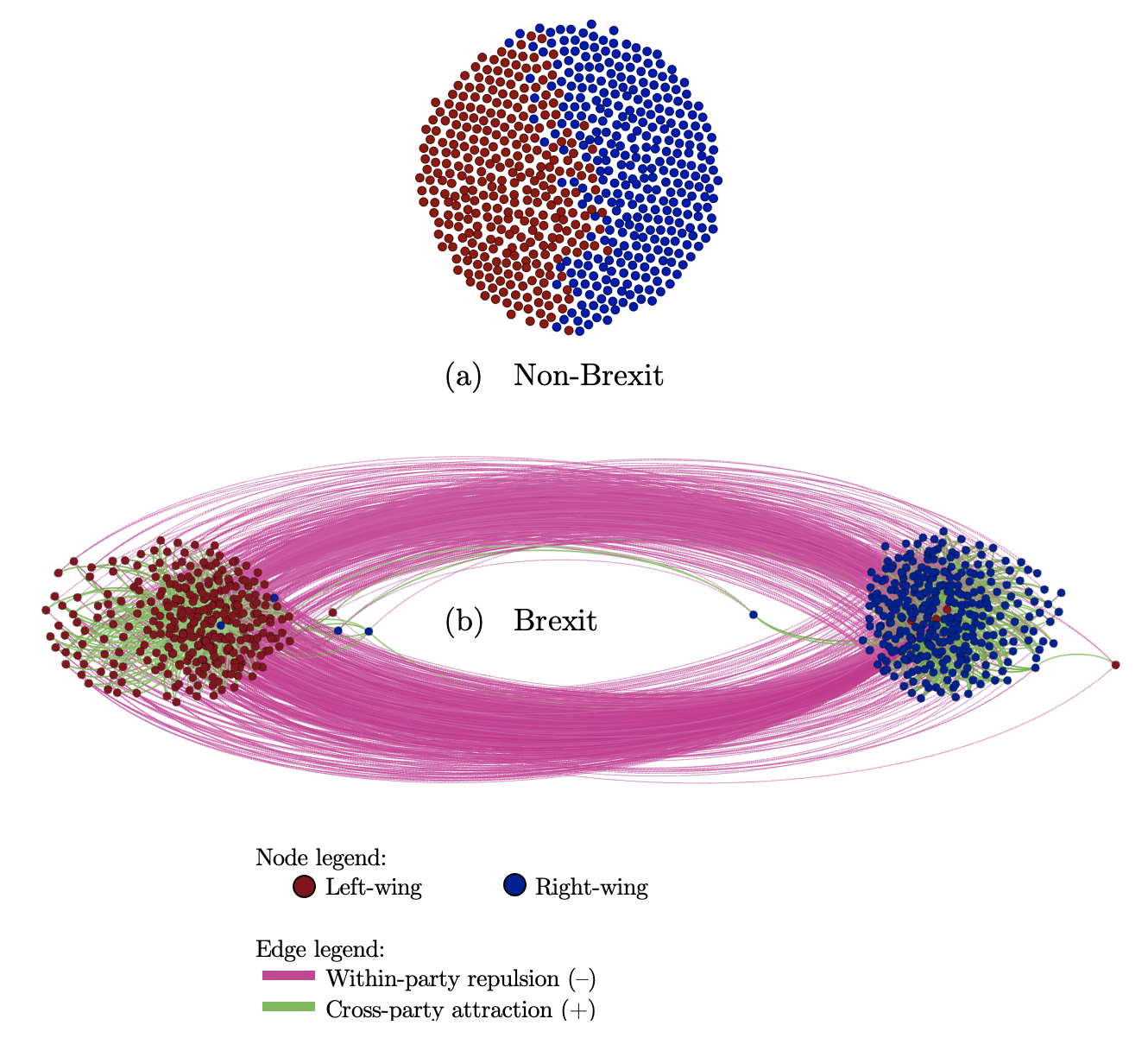}
\caption{Network projection of dissent and rebellion for non-Brexit (top) and Brexit divisions (bottom), for the 2 political ideology case. The absence of edges in the non-Brexit case suggest full ideological cohesion.}
\label{2p_network}
\end{figure}

Recall that in the right-side chart in Figure \ref{8pp} we show the histogram of a two-ideology case, where we re-labelled the Conservatives, the DUP, and one independent MP as right-wing; while the rest of the MPs were classified as left-wing. Re-labelling each MP by ideology rather than by political party gives a clearer and more striking result. In both the histogram in Figure \ref{8pp} and the network visualisation in Figure \ref{2p_network}, it appears that the cross-ideology alliances and within-ideology conflicts in the non-Brexit case completely disappear when re-labelling the MPs by left or right ideology. The absence of edges in the top graph of the non-Brexit case suggests full cohesion within each left-wing and right-wing cluster, respectively. In contrast, the bottom graph depicting the Brexit case continues to show strong cross-cluster interactions, suggesting instances of rebellion on both cross-ideology alliances and within-ideology conflicts.  

\newpage
\subsection*{Eigenvector centrality}
In Table \ref{ev}, we list the summary statistics for the eigenvector centrality measure. The low mean and median centrality scores indicate that MPs generally adhere to their expected party rhetoric.  Meanwhile, the top 10 rebels, identified below and listed with their party affiliations and expected ideology, have eigenvector centrality scores of $>$0.9.

\begin{table}[htbp]
\centering
\scalebox{0.9}{
\begin{tabular}{lrrrr}
\multicolumn{1}{l}{\textbf{MP}} & \multicolumn{1}{r}{\textbf{\begin{tabular}[r]{@{}r@{}}Party\\\scriptsize(as of Apr 2019)\end{tabular}}} & \multicolumn{1}{r}{\textbf{\begin{tabular}[r]{@{}r@{}}Ideology\\\scriptsize(inferred from party)\end{tabular}}} & \multicolumn{1}{r}{\textbf{\begin{tabular}[r]{@{}r@{}}Eigenvector\\ centrality\end{tabular}}} \\ \hline
\multicolumn{1}{l}{1.  Lady Hermon**} & Independent  & Left & \multicolumn{1}{r}{1.0}\\
\multicolumn{1}{l}{2.  Mr Kenneth Clarke*} & Conservative & Right& \multicolumn{1}{r}{1.0}\\
\multicolumn{1}{l}{3.  Frank Field*} & Independent & Left & \multicolumn{1}{r}{0.99}\\
\multicolumn{1}{l}{4.  Graham Stringer*} & Labour & Left & \multicolumn{1}{r}{0.98}\\
\multicolumn{1}{l}{5.  Kate Hoey*} & Labour & Left & \multicolumn{1}{r}{0.98}\\
\multicolumn{1}{l}{6.  Kelvin Hopkins*} & Independent & Left  & \multicolumn{1}{r}{0.97}\\
\multicolumn{1}{l}{7.  Anna Soubry*} & Conservative & Right & \multicolumn{1}{r}{0.96}\\
\multicolumn{1}{l}{8.  Dr Sarah Wollaston*} & Conservative  & Right & \multicolumn{1}{r}{0.92}\\
\multicolumn{1}{l}{9.  Heidi Allen*} & Conservative & Right & \multicolumn{1}{r}{0.91}\\
\multicolumn{1}{l}{10. Mr Ronnie Campbell} & Labour & Left & \multicolumn{1}{r}{0.91}\\
\hline
\multicolumn{3}{l}{   Mean centrality}  & 0.16 \\
\multicolumn{3}{l}{   Median centrality}  & 0.14\\
\multicolumn{3}{l}{   Number of MPs} & 639\\
\end{tabular}}
\caption{Top 10 rebels based on eigenvector centrality.  Asterisks (*) indicate that the MP was also identified by the network visualisation. Lady Hermon (**) was not identified a rebel by visual inspection, but was labelled an Independent rebel in our Data and Methods section.}
\label{ev}
\end{table}

Comparing the results of the eigenvector centrality metric with our 2-ideology network visualisation, we are able to correctly infer the top 9 of the 10 rebels. The visualisation is provided in Figure \ref{2p_network_labeled} below. Recall that in one case which we referred to in our Data section, we classified Lady Hermon (ranked number 1 rebel by eigenvector centrality) as a Left-wing Independent because of her refusal to affiliate with her former party, the Right-leaning Ulster Unionist Party.  The centrality score rightfully detected her as a rebel in spite of her network position of being in the Left-wing. Thus the results of the centrality measure can accurately quantify and identify dissenters regardless of arbitrary labeling (false positives), particularly in the case of Independent MPs' ideologies in the network visualisation. 

\begin{figure}[htbp]
\centering
\includegraphics[width=\textwidth]{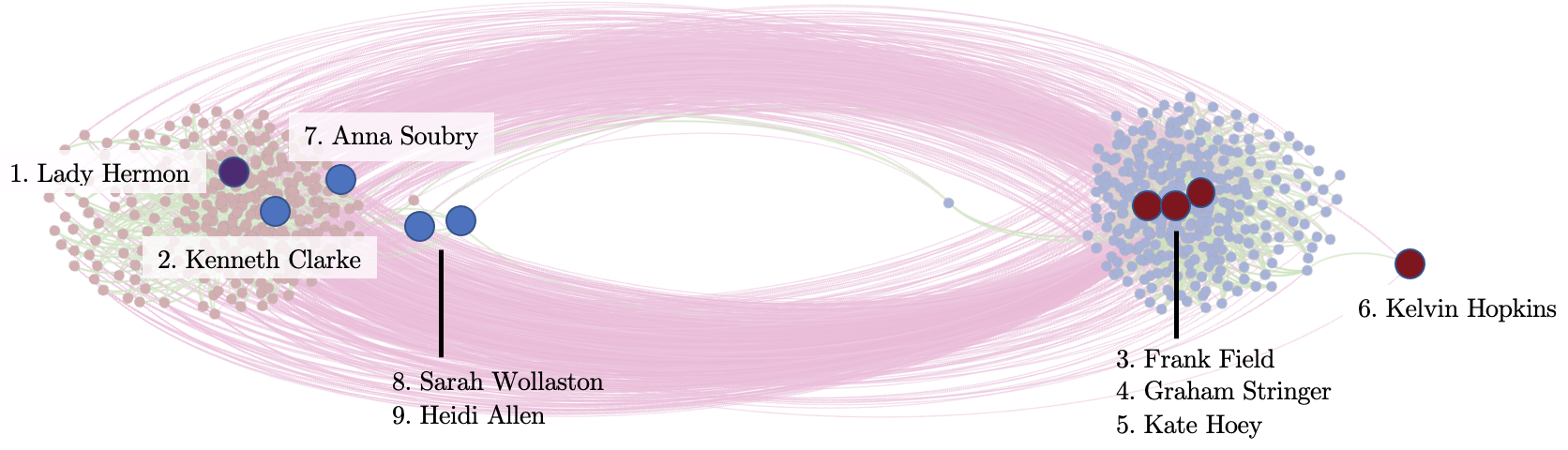}
\caption{The eigenvector centrality measure correctly identified the top 9 of 10 rebels. Note that Lady Hermon was labeled an Independent rebel in the Data and Methods section.} 
\label{2p_network_labeled}
\end{figure}

\subsection*{Validation: How MPs voted on BoJo's Brexit deal}
On October 22, 2019, after three failed attempts by Theresa May in getting a Brexit deal passed in Parliament, MPs have agreed on a way forward with Boris Johnson's Withdrawal Agreement Bill, backing his proposal to a vote of 329-299. However, while this was a win in principle, minutes later, MPs voted to reject the Brexit bill's accelerated timetable, 322-308.

To further validate our results, we investigated the voting behaviour of each MP on our list in Table \ref{ev}, and checked whether they rebelled against their party in voting for Boris Johnson's Brexit deal timetable. The vote tallies by MP are taken from the \citet{bojo_2019} and Hansard, and the comparative table is presented in Table \ref{rebelbojo} below.

Of the ten MPs in list on Table \ref{rebelbojo}, nine had cast a vote and eight of them rebelled against their party or ideological expectation in Table \ref{ev}. In the Supplementary Material we also provided an expanded version of this list, where we used the eigenvector centralities of all 639 MPs in order to calculate the overall accuracy score of our model. The accuracy score is defined by the number of correct predictions (true positives + true negatives) divided by total predictions, and the accuracy score of our model is 94\%.

In anticipating voting patterns as they unfold, these results provide further indication that the centrality metric and the findings from the network visualisation are valid. 

\begin{table}[htbp]
\centering
\scalebox{0.9}{
\begin{tabular}{lrrrr}
\multicolumn{1}{l}{\textbf{MP}} & \multicolumn{1}{r}{\textbf{\begin{tabular}[r]{@{}r@{}}Party\\\scriptsize(as of Apr 2019)\end{tabular}}} & \multicolumn{1}{r}{\textbf{\begin{tabular}[r]{@{}r@{}}Party's vote\\expectation\\\scriptsize(for/against Brexit bill)\end{tabular}}} & \multicolumn{1}{r}{\textbf{\begin{tabular}[r]{@{}r@{}}MP \\ voted*\end{tabular}}} \\ \hline
\multicolumn{1}{l}{Lady Hermon} & Independent  & For & Against \\
\multicolumn{1}{l}{Mr Kenneth Clarke} & Conservative & For & Against \\
\multicolumn{1}{l}{Frank Field} & Independent & Against & For \\
\multicolumn{1}{l}{Graham Stringer} & Labour & Against & Against \\
\multicolumn{1}{l}{Kate Hoey} & Labour & Against & For \\
\multicolumn{1}{l}{Kelvin Hopkins} & Independent & Against  & For  \\
\multicolumn{1}{l}{Anna Soubry} & Conservative & For & Against  \\
\multicolumn{1}{l}{Dr Sarah Wollaston} & Conservative  & For & Against  \\
\multicolumn{1}{l}{Heidi Allen} & Conservative & For & Against  \\
\multicolumn{1}{l}{Mr Ronnie Campbell} & Labour & Against & - \\
 \hline
\multicolumn{4}{l}{\begin{tabular}[c]{@{}l@{}}\scriptsize * Note: Ken Clarke voted to pass the Brexit bill but voted against the expedited timeline \\ \quad \quad \scriptsize Kate Hoey did not vote on the Brexit bill, but voted for the expedited timeline.\end{tabular}}
\end{tabular}}
\caption{Division votes on the Oct 2019 Brexit bill timeline of top 10 rebels compared to what is expected by their party. The accuracy score of the full model is 94\%, and calculations are provided in the Supplementary Material.}
\label{rebelbojo}
\end{table}

\bigskip
\newpage
\section*{Conclusion}
The study of dissent and rebellion is not a new concept, but through social network analysis we aim to bring new perspectives.  We contribute this alternative methodology of using pairwise interactions between MPs as an enhancement to the current cohesion metrics which are aggregated and binary in nature.  After all, the voting process is a social phenomenon, and our understanding of Parliamentary rebellion goes beyond knowing whether dissent levels are high or if cohesion levels within parties are low.  Especially in today's localism-focused politics, what adds value is our understanding of the dynamics behind every MP's vote, and anticipating voting patterns as they unfold.  When faced with uncertainty, learning that there are splits and factions in the party system is `stating the obvious' as it is already after-the-fact and it contributes very little information in the action plan to move forward.  But understanding the interactions of individuals on a granularised  (division) level can provide an ample road map to overcome real-time challenges in the legislative process.

We can count a few limitations in our approach. First, it is by no means causal.  While we can identify the rebels by visual inspection of their voting behaviour or by computing their rebellion scores, whether or not the MP will actually leave his/her party remains to be seen.  As a preliminary step, we used the validation section as a prelude to the causality discussion, but it is recommended that the social network analysis is supplemented by other quantitative approaches that specialise in causal inference. Second, by taking a granular approach, one may overlook the benefits of using a simpler and uncomplicated measure of party cohesion.  For instance, traditional aggregated approaches may be easier to interpret as it is computationally convenient and most information is condensed to a single number (e.g. a party cohesion score), whereas in producing the network, or a rebellion score for each MP may entail calculations and matrix transformations that may be difficult to disentangle.  In this analysis we tried to bridge this complexity gap by producing visual representations of the network, but the process of generating it may not necessarily be straightforward.

Our analysis suggests that European integration continues to linger until the present, and that euroscepticism still casts a long shadow in the House of Commons.  Our immediate experience does not give us much credence, as transition governments following a period of fragmentation are likely to be fundamentally weak.  But on the other hand, given today's empirical methods, we no longer are limited to using historical accounts to benchmark the future. The record-keeping of the UK Parliament is one of the most modern, and comprehensive systems available for legislative data.  Contemporary computational methods such as social network analysis may not be typically used in the political sciences, but the methodology lends a fresh take in understanding social phenomenon particularly in voting behaviour. The methodological framework suggested here can be used in future replication of the same analysis on the data from the forthcoming sessions of the Parliament in a continuous manner and the results of such analysis can inform the politicians, political analysts, and most importantly the citizens. 

\section*{Declarations}
\subsection*{Data Availability}
The data, replication instructions, and the data's codebook can be found at \url{https://doi.org/10.7910/DVN/OR05MA}.

\subsection*{Competing interests}
The authors declare that they have no competing interests.

\subsection*{Funding}
TY was partially supported by the Alan Turing Institute under the EPSRC grant no. EP/N510129/1.

\subsection*{Authors' contributions}
CI analyzed the data and was a major contributor in writing the manuscript. TY designed the study and the analysis, and contributed to the writing of the manuscript. Both authors read and approved the final manuscript.

\subsection*{Acknowledgement}
We would like to thank Rose de Geus for useful discussions.

\bibliographystyle{chicago}
\bibliography{biblio}

\begin{thebibliography}{}

\bibitem[\protect\citeauthoryear{Baker, Gamble, Ludlam, and Seawright}{Baker
  et~al.}{1999}]{baker1999backbench}
Baker, D., A.~Gamble, S.~Ludlam, and D.~Seawright (1999).
\newblock Backbenchers with attitude: a seismic study of the conservative party
  and dissent on europe.
\newblock {\em Party discipline and parliamentary government\/}, 72--93.

\bibitem[\protect\citeauthoryear{{BBC}}{{BBC}}{2010}]{bbc_mp_2010}
{BBC} (2010, March).
\newblock {MP} {Hermon} quits {Ulster} {Unionists}.
\newblock {\em British Broadcasting Corporation\/}.
\newblock
  \url{http://news.bbc.co.uk/2/hi/uk_news/northern_ireland/8586845.stm}. Last
  accessed 2019-07-03.

\bibitem[\protect\citeauthoryear{{BBC}}{{BBC}}{2019}]{bojo_2019}
{BBC} (2019, October).
\newblock Brexit bill `in limbo' as {MPs} reject timetable.
\newblock {\em BBC\/}.
\newblock \url{https://www.bbc.com/news/uk-politics-50146182}. Last accessed on
  2020-11-16.

\bibitem[\protect\citeauthoryear{Bonacich}{Bonacich}{2007}]{bonacich2007some}
Bonacich, P. (2007).
\newblock Some unique properties of eigenvector centrality.
\newblock {\em Social networks\/}~{\em 29\/}(4), 555--564.

\bibitem[\protect\citeauthoryear{Collie}{Collie}{1984}]{collie_voting_1984}
Collie, M.~P. (1984).
\newblock Voting {Behavior} in {Legislatures}.
\newblock {\em Legislative Studies Quarterly\/}~{\em 9\/}(1), 3--50.

\bibitem[\protect\citeauthoryear{Cook, Emerson, Gillmore, and Yamagishi}{Cook
  et~al.}{1983}]{cook1983}
Cook, K.~S., R.~M. Emerson, M.~R. Gillmore, and T.~Yamagishi (1983).
\newblock The distribution of power in exchange networks: Theory and
  experimental results.
\newblock {\em American Journal of Sociology\/}~{\em 89\/}(2), 275--305.
\newblock Cited By :493.

\bibitem[\protect\citeauthoryear{Cowley and Norton}{Cowley and
  Norton}{1999}]{cowley_rebels_1999}
Cowley, P. and P.~Norton (1999, April).
\newblock Rebels and rebellions: {Conservative} {MPs} in the 1992 {Parliament}.
\newblock {\em The British Journal of Politics \& International
  Relations\/}~{\em 1\/}(1), 84--105.

\bibitem[\protect\citeauthoryear{Forster}{Forster}{2002}]{forster_euroscepticism_2002}
Forster, A. (2002).
\newblock {\em Euroscepticism in contemporary {British} politics: opposition to
  {Europe} in the {British} {Conservative} and {Labour} parties since 1945}.
\newblock London: Routledge.

\bibitem[\protect\citeauthoryear{Garner and Letki}{Garner and
  Letki}{2005}]{garner_party_2005}
Garner, C. and N.~Letki (2005).
\newblock Party {Structure} and {Backbench} {Dissent} in the {Canadian} and
  {British} {Parliaments}.
\newblock {\em Canadian Journal of Political Science / Revue canadienne de
  science politique\/}~{\em 38\/}(2), 463--482.

\bibitem[\protect\citeauthoryear{Jacomy, Venturini, Heymann, and
  Bastian}{Jacomy et~al.}{2014}]{jacomy_forceatlas2_2012}
Jacomy, M., T.~Venturini, S.~Heymann, and M.~Bastian (2014, June).
\newblock {ForceAtlas}2, a {Continuous} {Graph} {Layout} {Algorithm} for
  {Handy} {Network} {Visualization} {Designed} for the {Gephi} {Software}.
\newblock {\em PLOS ONE\/}~{\em 9\/}(6), e98679.

\bibitem[\protect\citeauthoryear{Martin, Zhang, and Newman}{Martin
  et~al.}{2014}]{martin2014}
Martin, T., X.~Zhang, and M.~E.~J. Newman (2014, Nov).
\newblock Localization and centrality in networks.
\newblock {\em Phys. Rev. E\/}~{\em 90}, 052808.

\bibitem[\protect\citeauthoryear{Norton}{Norton}{1980}]{norton_changing_1980}
Norton, P. (1980).
\newblock The {Changing} {Face} of the {British} {House} of {Commons} in the
  1970s.
\newblock {\em Legislative Studies Quarterly\/}~{\em 5\/}(3), 333--357.

\bibitem[\protect\citeauthoryear{Pattie, Johnston, and Stuart}{Pattie
  et~al.}{1998}]{pattie1998}
Pattie, C., R.~Johnston, and M.~Stuart (1998).
\newblock Voting without party?
\newblock In P.~Cowley (Ed.), {\em Conscience and Parliament}, pp.\  162. Frank
  Cass, London.

\bibitem[\protect\citeauthoryear{Raymond and Worth}{Raymond and
  Worth}{2017}]{raymond_explaining_2017}
Raymond, C.~D. and R.~M. Worth (2017, November).
\newblock Explaining voting behaviour on free votes: {Solely} a matter of
  preference?
\newblock {\em British Politics\/}~{\em 12\/}(4), 555--564.

\bibitem[\protect\citeauthoryear{Rice}{Rice}{1938}]{rice1938}
Rice, S.~A. (1938).
\newblock Quantitative methods in politics.
\newblock {\em Journal of the American Statistical Association\/}~{\em
  33\/}(201), 126--130.

\bibitem[\protect\citeauthoryear{Russell}{Russell}{2014}]{russell_parliamentary_2014}
Russell, M. (2014, September).
\newblock Parliamentary party cohesion: {Some} explanations from psychology.
\newblock {\em Party Politics\/}~{\em 20\/}(5), 712--723.

\bibitem[\protect\citeauthoryear{Searing}{Searing}{1978}]{searing_measuring_1978}
Searing, D.~D. (1978).
\newblock Measuring {Politicians}' {Values}: {Administration} and {Assessment}
  of a {Ranking} {Technique} in the {British} {House} of {Commons}.
\newblock {\em The American Political Science Review\/}~{\em 72\/}(1), 65--79.

\bibitem[\protect\citeauthoryear{Sieberer}{Sieberer}{2006}]{sieberer_party_2006}
Sieberer, U. (2006, June).
\newblock Party unity in parliamentary democracies: {A} comparative analysis.
\newblock {\em The Journal of Legislative Studies\/}~{\em 12\/}(2), 150--178.

\bibitem[\protect\citeauthoryear{Surridge}{Surridge}{2018}]{surridge_brexit_2018}
Surridge, P. (2018, December).
\newblock Brexit, {British} {Politics}, and the {Left}-{Right} {Divide}.
\newblock {\em Political Insight\/}~{\em 9\/}(4), 4--7.

\bibitem[\protect\citeauthoryear{Tang, Chang, Aggarwal, and Liu}{Tang
  et~al.}{2016}]{tang2016survey}
Tang, J., Y.~Chang, C.~Aggarwal, and H.~Liu (2016).
\newblock A survey of signed network mining in social media.
\newblock {\em ACM Computing Surveys (CSUR)\/}~{\em 49\/}(3), 1--37.

\bibitem[\protect\citeauthoryear{Tzelgov}{Tzelgov}{2014}]{tzelgov_cross-cutting_2014}
Tzelgov, E. (2014, March).
\newblock Cross-cutting issues, intraparty dissent and party strategy: {The}
  issue of {European} integration in the {House} of {Commons}.
\newblock {\em European Union Politics\/}~{\em 15\/}(1), 3--23.

\bibitem[\protect\citeauthoryear{Vasilopoulou}{Vasilopoulou}{2016}]{vasilopoulou_uk_2016}
Vasilopoulou, S. (2016, April).
\newblock {UK} {Euroscepticism} and the {Brexit} {Referendum}.
\newblock {\em The Political Quarterly\/}~{\em 87\/}(2), 219--227.

\bibitem[\protect\citeauthoryear{Whiteley and Seyd}{Whiteley and
  Seyd}{1999}]{whiteley1999discipline}
Whiteley, P.~F. and P.~Seyd (1999).
\newblock Discipline in the british conservative party: The attitudes of party
  activists toward the role of their members of parliament.
\newblock pp.\  53--71. Ohio State University Press Columbus, OH.

\bibitem[\protect\citeauthoryear{Wood}{Wood}{1982}]{wood_comparing_1982}
Wood, D.~M. (1982).
\newblock Comparing {Parliamentary} {Voting} on {European} {Issues} in {France}
  and {Britain}.
\newblock {\em Legislative Studies Quarterly\/}~{\em 7\/}(1), 101--117.

\end{thebibliography}

\newpage
\section*{Supplementary Material}

\renewcommand\thefigure{S\arabic{figure}}    
\renewcommand\thetable{S\arabic{table}}    
\setcounter{figure}{0} 
\setcounter{table}{0} 

\subsection*{List of Abbreviations}
\begin{itemize}
\item[MP] Members of Parliament, or persons elected by all those who live in a particular area (constituency) to represent them in the House of Commons. MPs consider and propose new laws, and can scrutinise government policies by asking ministers questions about current issues either in the Commons Chamber or in Committees

\item[UK] United Kingdom of Great Britain and Northern Ireland

\item[NATO] North Atlantic Treaty Organization
\item[EU] European Union
\item[EEC] European Economic Commission
\item[BBC] British Broadcasting Corporation
\item[DUP] Democratic Unionist Party
\item[SNP] Scottish National Party
\end{itemize}

\subsection*{Hansard}

The following are the most relevant links related to the data extraction phase:

\begin{itemize}
    \item \href{http://explore.data.parliament.uk/?endpoint=commonsdivisions#download-list.}{http://explore.data.parliament.uk/?endpoint=commonsdivisions}\\ This contains the URI code, i.e. a six-digit identifier for each division: 1109556, 1108905, etc.
    \item \url{http://lda.data.parliament.uk/commonsdivisions/id/1109556.json}\\  This is the link which I use to extract the voting (Aye / No) data for each MP.  Note that I replace the 6-digit URI code to refer to different divisions.
\end{itemize}

\subsection*{Network Visualization}
The base framework of the force-directed algorithm that is used in this work is taken from \citet{jacomy_forceatlas2_2012}. The attraction force $F_a$ between two nodes $\mathbf {a}$ and $\mathbf {b}$ is directly related to its edge, or the weighted distance $w(e) \times d(\mathbf {a}, \mathbf {b})$:
$$F_a = w(e) \times d( \mathbf {a},\mathbf {b} )$$

Meanwhile, the repulsion force $F_d$ between two nodes $\mathbf {a}$ and $\mathbf {b}$ is a function of their weighted distance, each node's number of links (i.e. the degree) and a parameter $k_r$:
$$F_r ( \mathbf {a},\mathbf {b} )=k_r \frac{(deg(\mathbf {a})+1)(deg(\mathbf {b})+1)}{w(e) \times d( \mathbf {a},\mathbf {b} )}$$
$$ \text{where }k_r \text{ are user-defined settings on the gravity and scaling of the network } $$

The other details in the network implementation are as follows. For full definitions and explanations for each parameter, refer to \citet{jacomy_forceatlas2_2012}.

\begin{itemize}
    \item \textbf{Number of threads} imply more speed (more multithreading jobs).  The setting was set to 3.
    \item \textbf{Tolerance} implies the amount of swinging, and a lower number implies more precision.  The setting was 1 (default).
    \item \textbf{Scaling} is the repulsion parameter of the graph, where higher numbers show greater sparsity.  The setting was set to 2.
    \item \textbf{Gravity} attracts nodes to the center, and prevents nodes from drifting.  The setting was default (1).
    \item \textbf{Edge weight influence} was set to ``normal.'' The other option was ``no influence.''
\end{itemize}

\newpage
\subsection*{Accuracy score calculation}
Using the computed eigenvector centralities for each MP, we compared each predicted result with their actual vote on the Brexit bill proposed on October 22, 2019. There were two debates held that day on the withdrawal agreement bill, one at 7:00 pm, where the Ayes won 329-299; and a second debate at 7:16 pm on the expedited timeline, where the Noes won 308-322.  We compared their predicted result with their response to the second debate.

\begin{figure}[htbp]
\centering
\includegraphics[width=\textwidth]{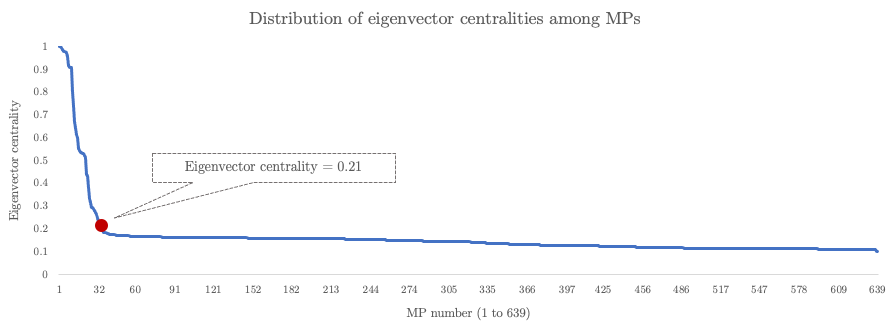}
\caption{The eigenvector distribution.} 
\label{ev_supl}
\end{figure}

We predicted the rebels using the elbow method, where we plotted the distribution of all 639 eigenvector centralities and selected the inflection point of the curve as the cut-off.  The elbow of the distribution is at the value 0.21, as illustrated in Figure \ref{ev_supl}, and MPs with centralities to the left of the curve ($\ge0.21$) were predicted as rebels.

Plotting the confusion matrix, we get the following result in Table \ref{cm}.  The accuracy score, calculated by $(TN+TP)/(TN+FP+TP+FN)$, is $(580+20)/(20+14+25+580)= 94\%$.

\begin{table}[htbp]
\begin{tabular}{llcc}
 & \multicolumn{1}{c}{} & \multicolumn{2}{c}{Actual} \\ \cline{3-4} 
 & \multicolumn{1}{l|}{} & \multicolumn{1}{l|}{Rebel = Yes} & \multicolumn{1}{l|}{Rebel = No} \\ \cline{2-4} 
\multicolumn{1}{c|}{{\begin{tabular}[c]{@{}c@{}}\\ Predicted\end{tabular}}} & \multicolumn{1}{l|}{Rebel = Yes} & \multicolumn{1}{c|}{20 (TP)} & \multicolumn{1}{c|}{14 (FP)} \\ \cline{2-4} 
\multicolumn{1}{c|}{} & \multicolumn{1}{l|}{Rebel = No} & \multicolumn{1}{c|}{25 (FN)} & \multicolumn{1}{c|}{580 (TN)} \\ \cline{2-4} 
\end{tabular}
\caption{Confusion matrix of voting outcomes for all 639 MPs.}
\label{cm}
\end{table}

\end{document}